\documentclass{epl}
\usepackage{amssymb}

\usepackage{graphicx}
\usepackage{amsmath}

\institute{Department of Materials and Interfaces, Weizmann Institute, Rehovot 76100,
Israel}

\begin{document}

\title{Universal reduction of pressure between charged surfaces by long-wavelength
surface charge modulation}
\author{D. B. Lukatsky and S. A. Safran}
\maketitle

\begin{abstract}
We predict theoretically that long-wavelength surface charge modulations
universally reduce the pressure between the charged surfaces with
counterions compared with the case of uniformly charged surfaces with the
same average surface charge density. The physical origin of this effect is
the fact that surface charge modulations \emph{always} lead to enhanced
counterion localization near the surfaces, and hence, fewer charges at the
midplane. We confirm the last prediction with Monte Carlo simulations.
\end{abstract}

\textit{Introduction-} The interaction of charged interfaces is a basic
problem in surface science, and is relevant to \textit{e.g.}, colloid
stabilization, membrane adhesion, microemulsion formation \cite
{Israelachvili}. While many theories have focused on the simple case of
homogeneously charged surfaces, real interfaces are characterized by
discrete surface charge distributions. Theoretical analysis of the problem
predicts two important properties of these systems: the spatial dependence
of the counterion/salt density and the pressure between the interfaces \cite
{old1,old2,old3,old4,old5,old6,old7,Andelman,Urbach,Fyl2002,Messina2001,NetzEPL2001a,DimaEPL2002,Hansen2002}%
. The main question is whether one can provide \emph{universal} predictions
of the effect of surface charge discreteness (modulations) on the counterion
spatial distribution and on the interaction law between the interfaces,
independently of the details of the surface distribution of discrete charges
(modulations). Our work shows that indeed, there are quite general effects
induced by surface charge discreteness.

The principal conclusion of previous studies based on the solution of linear
or quasi-linear Debye-Huckel equation with non-uniform surface charge
distributions with either excess salt \cite{old1,old2,Hansen2002},\ or with
counterions only \cite{Hansen2002},\ is the following: The discreteness (or
modulations) of surface charge \emph{always }leads to an \emph{enhanced
repulsive} interaction between the surfaces compared with the case of
uniformly charged surfaces with the same average surface charge density,
provided the charges (or charge modulations) on the two surfaces are
in-phase. If the charges (charge modulations) on the two surfaces are
out-of-phase, the repulsion can be \emph{either enhanced or reduced }%
compared with the uniformly charged surface case. These studies \cite
{old1,old2,Hansen2002} calculated the effects of charge modulations on the
counterion spatial distribution and on the electrostatic potential to first
order in the amplitude of modulations. The only relevant terms in the
pressure are therefore due to the square of the non-zero components of the
electric field, $\vec{E}$, at the midplane $\sim
\!E_{x}^{2}+E_{y}^{2}-E_{z}^{2}$ (see below). This contribution is always
positive for the in-phase case, and leads to the \textit{enhanced repulsion}
in the in-phase case.

In this paper, we show that a consistent calculation of the pressure,
including both the osmotic terms proportional to the counterion charge
density at the midplane, and the electric field terms calculated \textit{to
second order in the amplitude of modulations}, results in a\emph{\ universal
reduction} of the pressure by long-wavelength surface charge modulations.
The notion of a \emph{universal} reduction means that the effect does not
depend on a particular form, amplitude, and phase shift between surface
charge modulations. The physical origin of the predicted reduction of the
pressure is due to the osmotic effect and arises because surface charge
modulations \emph{always }enhance counterion localization in the vicinity of
surfaces, and hence, \emph{always }reduce the midplane counterion density.
We confirm this prediction for the charge density with Monte-Carlo (MC)
simulations.%
We emphasize that the predicted effects are due to the correlations between
the counterions and the inhomogeneities of the surface charge distribution.

\bigskip

\textit{Two surfaces with charge modulations- }In this section we consider
two, non-uniformly charged planar surfaces with fixed surface charge
densities $-e\sigma _{1}(\vec{\rho})$ and $-e\sigma _{2}(\vec{\rho})$ [where 
$\vec{\rho}=(x,y)$], respectively. The surfaces whose average charges are
equal and are taken to be negative, are located at $z=-\,\frac{h}{2}$ and $z=%
\frac{h}{2}$, respectively. The surface electrostatic repulsion is screened
by the positively charged counterions of valence $Z$; the counterions are
located only in the water between the surfaces, and the dielectric constant, 
$\epsilon $, ($\epsilon \simeq 80$ for water), is assumed to be homogeneous
throughout the system. The charge densities $\sigma _{1}(\vec{\rho})$ and $%
\sigma _{2}(\vec{\rho})$ obey the condition: $\int \sigma _{i}(\vec{\rho})\,d%
\vec{\rho}/A_{0}=\sigma _{0}$, where $A_{0}$ is the surface area, and $%
\sigma _{0}=Z\,N\,/\,2A_{0}$ is the average number of surface charge per
unit area, with $N$ being the total number of counterions in the system.

The system is fully specified within mean-field theory, by solving the
differential equation for the potential \cite{SamBook}, $\varphi (\vec{\rho}%
,z)$, with the boundary conditions implied by the delta function, spatially
dependent surface charge distributions $\sigma _{1}(\vec{\rho})$ and $\sigma
_{2}(\vec{\rho})$: 
\begin{equation}
\nabla ^{2}\phi \left( \rho ,z\right) =-4\pi \ell _{B}Z\left[
Z\,n_{0}\,e^{-\phi }-\delta (z+\frac{h}{2})\sigma _{1}(\vec{\rho})-\delta (z-%
\frac{h}{2})\sigma _{2}(\vec{\rho})\right] \,,\text{ \ }|z|\leq \frac{h}{2},
\label{PB1}
\end{equation}
and $\nabla ^{2}\phi \left( \vec{\rho},z\right) =4\pi \ell _{B}Z[\delta (z+%
\frac{h}{2})\sigma _{1}(\vec{\rho})+\delta (z-\frac{h}{2})\sigma _{2}(\vec{%
\rho})]$, for $|z|\geq \frac{h}{2}$, where there are no counterions \cite
{footnote1}. Here $\phi \left( \vec{\rho},z\right) =eZ\varphi (\vec{\rho}%
,z)/k_{B}T$ is the reduced electrostatic potential, $\ell _{B}=\frac{e^{2}}{%
\epsilon k_{B}T}$ is the Bjerrum length, and $\delta (z)$ is the Dirac delta
function. In what follows, we consider a general, modulated charge
distribution described by a Fourier decomposition \cite{DimaEPL2002}: $%
\sigma _{1}(\vec{\rho})=\sigma _{0}\left[ 1+\sum_{\vec{Q}\neq 0}\varepsilon (%
\vec{Q})\,e^{i\vec{Q}\,\cdot \vec{\rho}}\right] $ and $\,\sigma _{2}(\vec{%
\rho})=\sigma _{0}\left[ 1+\sum_{\vec{Q}\neq 0}\varepsilon (\vec{Q})\,e^{i%
\vec{Q}\,\cdot (\vec{\rho}+\vec{b})}\right] $, where $\vec{b}$ is a phase
shift of the modulations on the two surfaces. The coefficient $\varepsilon (%
\vec{Q})$ is the amplitude of the corresponding $\vec{Q}$ mode; for example,
on a square lattice, $\vec{Q}=2\pi \vec{m}/a$, and $\vec{m}=(m_{x},m_{y})$,
where $m_{x}$ and $m_{y}$ are integers, and $a$ is a lattice constant. For a
periodic lattice of surface charges, one must keep the infinite sum over
reciprocal lattice vectors, $\vec{Q}$, and set $\varepsilon (\vec{Q})=1$. In
analogy with the analysis performed earlier for the case of a single,
isolated surface \cite{DimaEPL2002}, we expand the potential, $\phi \left( 
\vec{\rho},z\right) $, in powers of the charge modulation amplitude, $%
\varepsilon (\vec{Q})$: $\phi \left( \vec{\rho},z\right) \approx \phi
_{0}\left( z\right) +\phi _{1}\left( \vec{\rho},z\right) +\phi _{2}\left( 
\vec{\rho},z\right) +...$, where $\phi _{0}\left( z\right) =2\ln \cos k_{0}z$
is the solution of the PB equation \cite{SamBook} with uniform surface
charge densities, $-e\sigma _{0}$. The parameter $k_{0}$ is determined by
the charge conservation condition \cite{SamBook}: $k_{0}\lambda _{0}\tan (%
\frac{k_{0}h}{2})=1$, where $\lambda _{0}=1/2\pi \ell _{B}Z\sigma _{0}$. Our
objective is to obtain perturbatively the corrections $\phi _{1}\left( \vec{%
\rho},z\right) $ [linear in $\varepsilon (\vec{Q})$] and $\phi _{2}\left( 
\vec{\rho},z\right) $ [quadratic in $\varepsilon (\vec{Q})$]. Substituting
the expansion of $\phi \left( \vec{\rho},z\right) $ into Eq. (\ref{PB1}),
and collecting terms linear in $\varepsilon (\vec{Q})$, we find that $\phi
_{1}(\vec{Q},z)$ [the two-dimensional Fourier transform of $\phi _{1}\left( 
\vec{\rho},z\right) $] obeys the following equation for $|z|\leq \frac{h}{2}$%
: 
\begin{equation}
\left[ -\,\frac{\partial ^{2}}{\partial z^{2}}+Q^{2}+\frac{2\,k_{0}^{2}}{%
\cos ^{2}k_{0}z}\right] \,\phi _{1}(\,\vec{Q},\,z)=-\,\frac{2\varepsilon (%
\vec{Q})}{\lambda _{0}}\left[ \delta (z+\frac{h}{2})+\delta (z-\frac{h}{2}%
)\,e^{i\,\vec{Q}\cdot \,\vec{b}}\right] ,\,\,\vec{Q}\neq 0.  \label{PB2}
\end{equation}

From here on, all the lengths in the problem will be rescaled in the units
of $\lambda _{0}$: $\mathcal{Q}\equiv Q\lambda _{0}$, $\tilde{z}\equiv
z/\lambda _{0}$, $\vec{\rho}\equiv \vec{\rho}/\lambda _{0}$, $k_{0}\equiv
k_{0}\lambda _{0}$, $h\equiv h/\lambda _{0}$, \textit{etc}. The solution of
this equation, $\phi _{1}(\,\mathcal{\vec{Q}},\,\tilde{z})$, is given in the 
\textit{Appendix}.

The equation for the contribution to the potential that is second order in
the surface charge modulation amplitude, $\phi _{2}(\vec{\rho},\tilde{z})$,
is obtained in a similar manner; collecting the terms quadratic in $%
\varepsilon (\vec{Q})$ we find for $|\tilde{z}|\leq \frac{h}{2}$: $\left[
-\nabla ^{2}+\frac{2\,k_{0}^{2}}{\cos ^{2}k_{0}\tilde{z}}\right] \,\phi _{2}(%
\vec{\rho},\tilde{z})=\frac{k_{0}^{2}}{\cos ^{2}k_{0}\tilde{z}}\,\phi
_{1}^{2}(\vec{\rho},\tilde{z})$. This equation is solved, using the Green
function for the homogeneous equation, \cite{Dima}. In what follows, for the
sake of simplicity, we consider two opposite cases: (i) in-phase, \textit{%
i.e.}, maximum interaction, $\vec{b}=(0,0)$, and (ii) out-of-phase,\textit{\
i.e.}, minimum interaction, $\vec{b}=(\frac{a}{2},\frac{a}{2})$. The case of
arbitrary $\vec{b}$ may be solved analytically, as well. We will consider
surface charge modulations described by a single symmetric $\mathcal{\vec{Q}}
$ mode: $\mathcal{\vec{Q}}=\{(\pm \frac{2\pi }{a},0),(0,\pm \frac{2\pi }{a}%
)\}$.

The average of $\phi _{2}$ in the $x\!\!\,-\!\!\,y$ plane is given in the 
\textit{Appendix} in both cases. Our main prediction is the \emph{enhancement%
} of the counterion density at each surface, averaged over the $%
x\!\!\,-\!\!\,y$ plane, $\left\langle n(\vec{\rho},\tilde{z}=\pm \frac{h}{2}%
)\right\rangle _{\vec{\rho}}$. The physical importance of this result is
that by charge conservation it leads to the \emph{reduction }of the midplane
average density compared to the case of uniformly charged surfaces in both
the in-phase, $\vec{b}=(0,0)$, and out-of-phase, $\vec{b}=(\frac{a}{2},\frac{%
a}{2})$, cases. This conclusion comes from the analysis of $\left\langle n(%
\vec{\rho},\tilde{z})\right\rangle _{\vec{\rho}}$ which takes the following
form to quadratic order in $\varepsilon (\mathcal{\vec{Q}})$: $\left\langle
n(\vec{\rho},\tilde{z})\right\rangle _{\vec{\rho}}=n_{0}(\tilde{z})[1+\frac{1%
}{2}\left\langle \phi _{1}^{2}(\vec{\rho},\tilde{z})\right\rangle _{\vec{\rho%
}}-\left\langle \phi _{2}(\vec{\rho},\tilde{z})\right\rangle _{\vec{\rho}}]$%
, where $n_{0}(\tilde{z})=n_{0}/\cos ^{2}k_{0}\tilde{z}$ is the counterion
density between uniformly charged surfaces \cite{SamBook}, $k_{0}^{2}=2\pi
\ell _{B}Z^{2}n_{0}$\textit{.} We emphasize that the first-order term, $\phi
_{1}(\vec{\rho},\tilde{z})$, vanishes after averaging with respect to $\vec{%
\rho}$, and that only quadratic terms in $\varepsilon (\mathcal{\vec{Q}})$
contribute to the average density; this observation is crucial for the
correct calculation of the osmotic pressure, as we show below. The
enhancement of the counterion density in the vicinity of the surfaces [and
the consequent reduction of the midplane density] is due to the correlations
between the counterions and the inhomogeneities of the surface charge
distribution. In particular, in the asymptotic limit of small inter-surface
separation, $h\ll 1$, $\mathcal{Q}h\ll 1$ the normalized average midplane
counterion density has the form: (i) in the in-phase case, $\frac{%
\left\langle n(\vec{\rho},0)\right\rangle _{\vec{\rho}}}{n_{0}(0)}\approx 1-%
\frac{2}{3}\sum_{\mathcal{\vec{Q}}}\frac{|\varepsilon (\mathcal{\vec{Q}}%
)|^{2}}{(2+\mathcal{Q})^{2}}\,h+\mathcal{O}(h^{2})$; (ii) in the
out-of-phase case,$\frac{\left\langle n(\vec{\rho},0)\right\rangle _{\vec{%
\rho}}}{n_{0}(0)}\approx 1-\frac{1}{6}\sum_{\mathcal{\vec{Q}}}|\varepsilon (%
\mathcal{\vec{Q}})|^{2}\,h^{2}+\mathcal{O}(h^{3})$. We stress that the
strongest \emph{reduction} of the midplane counterion density [and hence,
the strongest \emph{enhancement} of the contact density] occurs for the
long-wavelength modulations. This is in a qualitative agreement with the
predictions obtained for a single, isolated surface \cite{DimaEPL2002}. In
the opposite limit of large separations, $h\gg 1$, $\mathcal{Q}h\gg 1$, the
midplane density is reduced with an asymptotic form that is identical in
both limits, (i) and (ii): $\frac{\left\langle n(\vec{\rho},0)\right\rangle
_{\vec{\rho}}}{n_{0}(0)}\approx 1-\sum_{\mathcal{\vec{Q}}}\frac{%
4\,|\varepsilon (\mathcal{\vec{Q}})|^{2}\,(1+\mathcal{Q})}{(1+2\mathcal{Q}+2%
\mathcal{Q}^{2})^{2}}\frac{1}{h}+\mathcal{O}(\frac{1}{h^{2}})$.

\begin{figure}[tbp]
\caption{Normalized counterion density in the in-phase case, $\tilde{n}(%
\tilde{z})\equiv 2\protect\pi \ell _{B}Z^{2}\protect\lambda
_{0}^{2}\,\left\langle n(\vec{\protect\rho},\tilde{z})\right\rangle _{\vec{%
\protect\rho}}$ from the MC simulations (symbols) and the corresponding
predictions of the analytic, perturbation theory (curves). The parameters
used are:\ $g\equiv \frac{\ell _{B}Z^{2}}{\protect\lambda _{0}}=0.5$, $%
\protect\varepsilon _{0}=1$, $\mathcal{Q}=1$ (diamonds and dashed line). The
counterion profile between uniformly charged surfaces with the same average
surface densities of charge (stars and solid line) is shown for comparison.
The simulations were performed with 100 particles. Inset: The pressure
difference diagram, $p-p_{0}=0$, as a function of the rescaled wavelength of
modulations, $\mathcal{Q}$, and the inter-surface separation, $h$, in the
in-phase case, where $p_{0}$ is the pressure between uniformly charged
surfaces. We stress that already at $h\gtrsim 1.25$, charge modulations
reduce the pressure for any values of $\mathcal{Q}$.}
\label{Fig.1}\onefigure{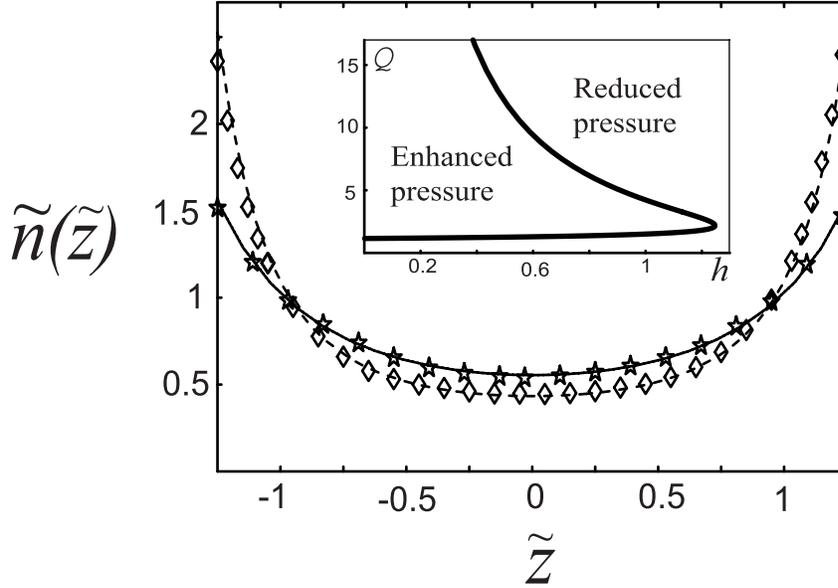}
\end{figure}

To verify the predictions of the theory, we have performed MC simulations of
the counterion density profiles between non-uniformly charged surfaces. The
non-uniform surface charge densities were chosen to have the form: $\sigma
_{1}(\vec{\rho})/\sigma _{0}=1+\varepsilon _{0}\sum_{\mathcal{\vec{Q}}}\,e^{i%
\mathcal{\vec{Q}}\,\cdot \vec{\rho}}$, and $\sigma _{2}(\vec{\rho})/\sigma
_{0}=1+\varepsilon _{0}\sum_{\mathcal{\vec{Q}}}\,e^{i\mathcal{\vec{Q}}%
\,\cdot (\vec{\rho}+\vec{b})}$, respectively, where $\mathcal{\vec{Q}}$
takes on the four symmetric values: $\{(\pm \frac{2\pi }{a},0),(0,\pm \frac{%
2\pi }{a})\}$. The counterions interact via an unscreened Coulomb potential,
and an Ewald 2D algorithm \cite{Simulations1} was used to sum the Coulomb
interactions; each counterion interacts with the surfaces via the potential, 
$eZ\varphi _{s}$\ generated by the \emph{exact} solution of the Poisson
equation with the chosen surface charge densities: $\frac{eZ\varphi _{s}}{%
k_{B}T}=-\varepsilon _{0}\sum_{\mathcal{\vec{Q}}}e^{i\mathcal{\vec{Q}}%
\,\cdot \vec{\rho}}(e^{-\mathcal{Q}|\tilde{z}+h/2|}+\,e^{i\mathcal{\vec{Q}}%
\,\cdot \vec{b}}e^{-\mathcal{Q}|\tilde{z}-h/2|})/\mathcal{Q}$. Using the
simulations we compare the counterion density averaged in $x\!-\!y$ plane.
We plot the normalized dimensionless density, $\tilde{n}(\tilde{z})\equiv
2\pi \ell _{B}Z^{2}\lambda _{0}^{2}\,\left\langle n(\vec{\rho},\tilde{z}%
)\right\rangle _{\vec{\rho}}$ vs. the rescaled length, $\tilde{z}\equiv
z/\lambda _{0}$. The analytic expressions for $\left\langle n(\vec{\rho},%
\tilde{z})\right\rangle _{\vec{\rho}}$ are given above (see also the \textit{%
Appendix}) for both the in-phase and the out-of-phase\ cases.

The dimensionless parameter which determines the strength of the Coulomb
interaction in the simulations is $g\equiv \frac{\ell _{B}Z^{2}}{\lambda _{0}%
}$; other two important parameters are the amplitude of the surface charge
modulation, $\varepsilon _{0}$ and the lengthscale, $\mathcal{Q}$, of the
surface charge modulations. In Fig. 1 we plot the simulated, average
density, $\tilde{n}(\tilde{z})$, in the in-phase case, $\vec{b}=(0,0)$ for
values of $g=0.5$, $\varepsilon _{0}=1$, and $\mathcal{Q}=1$ (diamonds), as
well as the simulated counterion profile between uniformly charged surfaces
with the same average surface charge density (stars). Additional simulations
showed that the average density is higher for the smaller values of $%
\mathcal{Q}$, similar to the conclusion obtained earlier in the case of a
single, isolated surface \cite{DimaEPL2002}. We also note that the
quantitative difference between average density profiles predicted by theory
and confirmed by simulations in the in-phase and the out-of-phase cases is
very small. The contact (midplane) density is always slightly higher
(smaller) in the in-phase case compared with the out-of-phase case, since in
the out-of-phase case the broken planar symmetry leads to the existence of
an oscillating $z-$ component of the electric field at the midplane (absent
in the in-phase case), that attracts more counterions to the midplane due to
the enhanced correlations. We stress again that the perturbation theory
prediction of the enhancement of the contact counterion density is universal
in the sense that the enhancement occurs for all values of the parameters, 
\textit{i.e.}, $\mathcal{Q}$, $h$, and $\varepsilon $, in both the in-phase
and the out-of-phase cases. We emphasize that ion fluctuation and
correlation effects neglected in the mean-field approach we adopt, become
important at high values of $g\gg 1$, \textit{i.e.}, in the strong coupling
regime \cite{NetzEPL2001a}. Finally we note, that qualitatively similar
conclusions with respect to the enhancement of the contact counterion
density have recently been obtained by numerical solutions of the integral
equation theory \cite{Fyl2002}; however some of their effects are related to
the fact that both the counterions and the surface charges have an excluded
volume in \cite{Fyl2002}.

\bigskip

\textit{Pressure between charge modulated surfaces-} In this section, we
show why surface charge modulations in the in-phase case (under certain
conditions), and in the out-of-phase-case (for all values of the parameters) 
\textit{always reduce }the\textit{\ }repulsive pressure between surfaces
with counterions compared with the case of uniformly charged surfaces.

The free energy, $F$, of the system has the form \cite{SamBook}: $\frac{F}{%
k_{B}T}=\int n(\vec{r}\,)[\ln (n(\vec{r}\,)v_{0})-1]\,d\vec{r}+\frac{1}{2}%
\int \phi (\vec{r}\,)n_{tot}(\vec{r}\,)\,d\vec{r}$, where the total system
charge density: $n_{tot}(\vec{r}\,)=n(\vec{r}\,)-\frac{1}{Z}[\sigma _{1}(%
\vec{\rho}\,)\delta (z+\frac{h}{2})+\sigma _{2}(\vec{\rho}\,)\delta (z-\frac{%
h}{2})]$, $v_{0}$ is an ion-size volume, and $\vec{r}=(\vec{\rho},z)$.
Keeping terms up to second order in the amplitude of surface charge
modulations, $\varepsilon (\vec{Q}\,)$, one notes that $F$ can be expressed
only in terms of the $x\!-\!y$ average of the potentials as obtained above, $%
\left\langle \phi _{2}(\vec{r}\,)\right\rangle _{\vec{\rho}}\,\,$, $%
\left\langle \phi _{1}^{2}(\vec{r}\,)\right\rangle _{\vec{\rho}}$ and $%
\left\langle \sigma _{i}(\vec{\rho}\,)\phi _{1}(\vec{r}\,)\right\rangle _{%
\vec{\rho}}$; after straightforward but lengthy integrations, one obtains
the following dimensionless free energy, $f\equiv \frac{F}{%
k_{B}TA_{0}(\sigma _{0}/Z)}$, per counterion $f_{+}$ and $f_{-}$ in the\emph{%
\ }in-phase and in the out-of-phase case, respectively: 
\begin{equation}
f_{\pm }=f_{0}+\sum_{\mathcal{\vec{Q}}}2|\varepsilon (\mathcal{\vec{Q}}%
)|^{2}\,\frac{1+\mathcal{Q}\pm e^{-\mathcal{Q}h}\,(\mathcal{Q}-1)}{1+2%
\mathcal{Q}+2\mathcal{Q}^{2}\mp e^{-\mathcal{Q}h}+(1\mp e^{-\mathcal{Q}%
h})\,k_{0}^{2}},  \label{free_energy}
\end{equation}
where $f_{0}$ is the free energy of uniformly charged surfaces with
counterions \cite{SamBook}. The normalized, dimensionless osmotic pressure
between the surfaces, $p=-\partial f/\partial h$, is the main quantity of
interest measured in experiments. Our principle observations follow
intuitively from the asymptotic analysis of the pressure:\ (i) in the\textit{%
\ in-phase} case, in the limit of small separations, $h\ll 1$, $\mathcal{Q}%
h\ll 1$, one has, 
\begin{equation}
p-p_{0}=\sum_{\mathcal{\vec{Q}}}\,|\varepsilon (\mathcal{\vec{Q}})|^{2}\,%
\frac{\mathcal{Q}^{2}-\frac{4}{3}}{(2+\mathcal{Q})^{2}}+\mathcal{O}(h),
\label{pressure_in}
\end{equation}
where $p_{0}=k_{0}^{2}$ is the normalized, dimensionless pressure \cite
{footnote2} between uniformly charged surfaces \cite{SamBook}. In the 
\textit{out-of-phase} case, when $h\ll 1$, $\mathcal{Q}h\ll 1$ we obtain, $%
p-p_{0}=-\sum_{\,\mathcal{\vec{Q}}}\,|\varepsilon (\mathcal{\vec{Q}})|^{2}+%
\mathcal{O}(h)$. Therefore in this asymptotic limit of small inter-surface
separations, in the in-phase case the pressure between the surfaces is
reduced for long-wavelength surface charge modulations; in the out-of-phase
case, the pressure is reduced for any wavelength. This effect [quadratic in $%
\varepsilon (\mathcal{\vec{Q}})$] is related to the reduction of the
midplane counterion density induced by surface charge modulations upon
decreasing of $\mathcal{Q}$.\ In the limit of large inter-surface
separations, $h\gg 1$, $\mathcal{Q}h\gg 1$, in both cases, (i) and (ii), the
pressure is\emph{\ always reduced}: 
\begin{equation}
p-p_{0}=-\sum_{\mathcal{\vec{Q}}}\,\frac{4|\varepsilon (\mathcal{\vec{Q}}%
)|^{2}\,(1+\mathcal{Q)}}{(1+2\mathcal{Q}+2\mathcal{Q}^{2})^{2}}\,\frac{\pi
^{2}}{h^{3}}+\mathcal{O}(\frac{1}{h^{4}}).  \label{pressure_long}
\end{equation}
We note that this asymptotic form is similar to the universal, Casimir-type
attractive fluctuation pressure, $\sim -1/h^{3}$ (see \textit{e.g.}, \cite
{SamEPL98,Kardar}) between the surfaces with boundary conditions that
suppress or modify fluctuations of the medium; the amplitude of this
pressure is non-universal, of course, since the spectrum of modes is
restricted by the fixed surface charge modulations in the present case.

It is interesting to note, that corresponding asymptotic results for the
pressure \cite{footnote2} between \textit{uniformly} charged surfaces \cite
{SamBook}, $p_{0}$ [ $p_{0}=\frac{2}{h}$, if $h\ll 1$, and $p_{0}=\frac{\pi
^{2}}{h^{2}}$, if $h\gg 1$] show that the modulation correction, $p-p_{0}$,
can \textit{never} overcome the overall repulsion in these limiting cases.
For intermediate values of $h$, however, the total pressure, $p$, can change
sign and become attractive, even in the in-phase case, if we extrapolate our
perturbative results to values of $\varepsilon (\mathcal{\vec{Q}})$, which
are not necessarily small. We also emphasize again that we use a mean-field
approach, and neglect the \textit{fluctuation }contribution to the pressure 
\cite{SamEPL98} that further reduces the repulsive interaction.

The generic behavior of the pressure difference, $p-p_{0}$, in the in-phase
case for arbitrary $h$ and $\mathcal{Q}$ is shown in the inset of Fig. 1.
There is a broad range of parameters where the repulsion between two
surfaces is reduced by the effect of surface charge modulations. It is
interesting to note, that already at $h\gtrsim 1.25$, charge modulations 
\emph{reduce }the pressure for any value of $\mathcal{Q}$. We have checked
that a qualitatively similar reduction of the repulsive pressure occurs in
the case of a square lattice of discrete charges (infinite sum with respect
to all $\mathcal{\vec{Q}}$ modes).

Our prediction of the \textit{reduction} of pressure in the in-phase case
differs from the conclusions of Refs. \cite{old1,old2,Hansen2002}.\ The
difference between those works and our predictions can be understood using
an alternative definition of the pressure via the stress tensor \cite{Russel}%
: $p=[\tilde{n}+\frac{1}{4}\left\langle \phi _{x}^{2}+\phi _{y}^{2}-\phi _{%
\tilde{z}}^{2}\right\rangle _{\vec{\rho}}]_{\tilde{z}=0}\,$, where $\tilde{n}%
\equiv 2\pi \ell _{B}Z^{2}\lambda _{0}^{2}\,\left\langle n\right\rangle _{%
\vec{\rho}}$ is the normalized, dimensionless density \cite{footnote2}; this
gives results identical to the ones presented above. In the in-phase case, $%
\phi _{\tilde{z}}=0$, as implied by the symmetry, and the electrostatic
contribution to the pressure is purely repulsive, arising from terms in the
electric field that are\emph{\ first order} in the amplitude of surface
charge modulations, $\varepsilon (\mathcal{\vec{Q}})$ [the pressure is, of
course, always quadratic in $\varepsilon (\mathcal{\vec{Q}})$, since $p\sim
\phi _{x}^{2}+\phi _{y}^{2}$]. This is the only contribution that was taken
into account in Refs. \cite{old1,old2,Hansen2002}, where only first order
corrections to the counterion density were calculated. As far as the osmotic
term is concerned, the first order corrections vanish when the average over
the $x\!-\!y$ plane is performed, and hence, it is necessary to calculate
the counterion density to \emph{second order} in the surface charge
modulation amplitude, $\varepsilon (\mathcal{\vec{Q}})$. It is this osmotic
contribution, calculated here, that is the origin\emph{\ }of the\emph{\
universal reduction} of the pressure, since $\tilde{n}$ at the midplane\ is
always reduced by surface charge modulations, as we have shown; this effect
is quadratic in $\varepsilon (\mathcal{\vec{Q}})$, and beyond the linear
approach of Refs. \cite{old1,old2,Hansen2002}. It is the competition between
the osmotic and electrostatic contribution that leads to the diagram
represented in the inset of Fig. 1. In the out-of-phase case, where $\phi _{%
\tilde{z}}\neq 0$, the electrostatic contribution to$\ p$ is reduced
compared with the in-phase case, and further reduction of the midplane
density by charge modulations \emph{always }leads to the reduction of the
pressure, compared with the case of uniformly charged surfaces.

In summary, we emphasize that our main result, the prediction of a \textit{%
universal} reduction of the osmotic pressure by long-wavelength, quenched
surface charge modulations, is naturally related to effects of fluctuating
surface charges on the pressure. The unifying theme is that systems with
thermally \textit{fluctuating} surface charge density are qualitatively
similar to systems with quenched, \textit{modulated} surface charge density:
in both situations, the counterions are more localized near the surfaces
compared with the case of uniformly charged surfaces, and in both cases this
leads to a reduction of the pressure.

\acknowledgements We are grateful to P. Pincus, C. Jeppesen, J.-P. Hansen,
A.W.C. Lau, R. Netz, and T.O. White. We acknowledge the support from Israel
Binational Science Foundation (BSF) grant 98-00063, and the support of the
Schmidt Minerva Center.

\bigskip

\textit{Appendix- }In this appendix, we summarize the results for
perturbative corrections, $\phi _{1}(\vec{\rho},\tilde{z})$, and $%
\left\langle \phi _{2}(\vec{\rho},\tilde{z})\right\rangle _{\vec{\rho}}$, to
the potential, $\phi _{0}(\tilde{z})$, induced by surface charge modulations.

In particular, one finds: $\phi _{1}(\,\mathcal{\vec{Q}},\,\tilde{z})=%
\mathcal{A}\,v_{1}(\mathcal{Q},\tilde{z})+\mathcal{B}\,v_{2}(\mathcal{Q},%
\tilde{z})$, where $\mathcal{A}=\frac{2\varepsilon (\mathcal{Q})}{\mathcal{Q}%
}\frac{\alpha _{1}+\alpha _{2}e^{i\,\mathcal{\vec{Q}}\cdot \,\vec{b}}}{%
\alpha _{1}^{2}-\alpha _{2}^{2}}$, $\mathcal{B}=\frac{2\varepsilon (\mathcal{%
Q})}{\mathcal{Q}}\frac{\alpha _{1}e^{i\,\mathcal{\vec{Q}}\cdot \,\vec{b}%
}+\alpha _{2}}{\alpha _{1}^{2}-\alpha _{2}^{2}}$, $\alpha _{1}=\exp (-\,\,%
\frac{\mathcal{Q}h}{2})(k_{0}^{2}+1)/\mathcal{Q}^{2}$, $\alpha _{2}=\exp (%
\frac{\mathcal{Q}h}{2})(k_{0}^{2}+1+2\mathcal{Q}+2\mathcal{Q}^{2})/\mathcal{Q%
}^{2}$, and $v_{1}(\mathcal{Q},\tilde{z})=e^{\mathcal{Q}\tilde{z}}(1+(k_{0}/%
\mathcal{Q})\,\tan k_{0}\tilde{z})$, $v_{2}(\mathcal{Q},\tilde{z})=e^{-%
\mathcal{Q}\tilde{z}}(1-(k_{0}/\mathcal{Q})\,\tan k_{0}\tilde{z})$.

The average potential, $\left\langle \phi _{2}(\vec{\rho},\tilde{z}%
)\right\rangle _{\vec{\rho}}$ is also obtained in a simple, closed form in
terms of elementary functions: $\left\langle \phi _{2}(\vec{\rho},\tilde{z}%
)\right\rangle _{\vec{\rho}}=\!\sum_{\mathcal{\vec{Q}}}\frac{16|\varepsilon (%
\mathcal{\vec{Q}})|^{2}\,\Gamma _{\pm }(\mathcal{Q})}{\mathcal{Q}%
^{2}\,(\alpha _{1}\mp \alpha _{2})^{2}}$, where $\Gamma _{+}(\mathcal{Q})$
and $\Gamma _{-}(\mathcal{Q})$ corresponds to the in-phase case and the
out-of-phase case, respectively. Here we use the notations: $\Gamma _{\pm }(%
\mathcal{Q})=u_{2}(\tilde{z})[\frac{1}{\eta }\mathcal{G}_{\pm }(\frac{k_{0}h%
}{2})-\mathcal{F}_{\pm }(\frac{k_{0}h}{2})+\mathcal{F}_{\pm }(k_{0}\tilde{z}%
)]-u_{1}(\tilde{z})\mathcal{G}_{\pm }(k_{0}\tilde{z})$, $u_{1}(\tilde{z}%
)=\tan k_{0}\tilde{z}$, $u_{2}(\tilde{z})=1+k_{0}\tilde{z}\tan k_{0}\tilde{z}
$, $\eta =\frac{k_{0}}{1+k_{0}^{2}}+\frac{k_{0}h}{2}$, $\mathcal{G}_{\pm
}(x)=\frac{\cosh 2\theta x\mp 1}{8\,\theta ^{2}\cos ^{2}x}\,[\tan x-x(1-\tan
^{2}x)]+\frac{\sinh 2\theta x}{4\theta \cos ^{2}x}\left[ 1+x\tan x\right]
\pm \frac{1}{4}\left[ \,\frac{x}{\cos ^{2}x}+\tan x\right] $, and $\mathcal{F%
}_{\pm }(x)=\frac{[\cosh 2\theta x\mp 1]\,\,\tan ^{4}x}{8\,\theta ^{2}}+%
\frac{\sinh 2\theta x\,\tan x\,}{4\theta \cos ^{2}x}\pm \frac{1}{4\cos ^{2}x}%
-\frac{\cosh 2\theta x}{8\,\theta ^{2}}$, where $\theta =\mathcal{Q}/k_{0}$.


\begin{thebibliography}{99}
\bibitem{footnote1}  The case where the counterions are located only in the
\ space between the surfaces is chosen for convenience to enable comparison
with MC simulations. The case where the counterions are located in the whole
space leads to similar qualitative conclusions.
\bibitem{footnote2}  We note again that the normalized, dimensionless
pressure, $p$, is related to the physical pressure, $\Pi $, by: $p\equiv 
\frac{\Pi }{k_{B}T\,2\pi \ell _{B}\sigma _{0}^{2}}$. In the case of
uniformly charged surfaces \cite{SamBook}, $\Pi _{0}=k_{B}T\,n_{0}$, and,
consequently, $p_{0}=k_{0}^{2}$, where $k_{0}$ [defined in the text after
Eq. (\ref{PB1})] is dimensionless [since we express the lengths in units of $%
\lambda _{0}$, as specified in the text after Eq. (\ref{PB2})].
\end{thebibliography}
\end{document}